# Risk Models as Mediating Artifacts: A Postphenomenological Analysis of the CIIM Framework in Cybersecurity Practice

Rommel Salas-Guerra, PhD

*Professor and researcher in cybersecurity*

*Professional Studies Program, Universidad Ana G. Méndez*

Salasc1@uagm.edu | https://ciim.drsalas.us

## Summary

This article applies postphenomenological theory to the field of cybersecurity risk management, arguing that formal risk models function as mediating artifacts that shape how security practitioners or analysts perceive, interpret, and act on threats. Based on Don Ihde's taxonomy on human-technology relationships and Peter-Paul Verbeek's extended mediational framework, the Contextual and Multimodal Hazard Impact Index (CIIM), an original dynamic risk model presented as an empirical case study, is analyzed. CIIM is formally defined as $CIIM(t+1) = [A \cdot T(t) \cdot V(t) \cdot E(t)] / R(t) + \alpha \cdot P(t)$, where the condition $R(t) \to 0$ is not treated as a computational artifact to be smoothed out, but as a genuine systemic collapse that signals singularity. This design choice constitutes a deliberate phenomenological move, allowing organizational fragility to be made visible in a way that previous CVSS-based and probabilistic models conceal. In addition, we examine how CIIM's time projection (t+1) and its hybrid machine learning architecture, combining LSTM/GRU, XGBoost, and Reinforcement Learning, produce a new form of technological intentionality that structures practitioner or analyst attention and ethical deliberation. The article concludes by establishing implications for the ethical design of cybersecurity instrumentation and for the post-phenomenological methodology itself, proposing the concept of 'phenomenology of collapse' as a contribution to the empirical philosophy of technology.

## 1. Introduction: The Invisible Architecture of Threat Perception

In different organizations, when a cybersecurity analyst opens a dashboard that shows a CVSS score of 9.8, what exactly do they see? The conventional answer taken from a positivist philosophy of science would state that it sees an objective measure of the severity of vulnerability, a number derived from a standardized and neutral formula. This article argues that this answer, although not incorrect, is profoundly incomplete, this is because the analyst does not simply see a number; He sees a world already structured by a particular technological artifact that has preformatted his perceptual field, having prioritized certain dimensions of risk over others, making certain courses of action seem natural and others invisible.

This statement is the starting point of postphenomenological research, as conceived by Don Ihde [1,2] and later developed by Peter-Paul Verbeek [4,5], Yoni Van Den Eede [23] and others, which leads us to understand that postphenomenology is not a critique of technology; on the contrary, it is an empirical philosophical methodology that examines, on a case-by-case basis, how specific technological artifacts examine the interaction between human beings and their world. Whereas Heidegger [7] established technology as a totalizing mode of revelation that reduces everything to a 'permanent reserve' (Bestand), postphenomenology therefore proposes a more efficient approach based on how different technologies mediate differently; Understanding these differences requires careful attention to the materiality and design of artifacts.

Today, the philosophy of cybersecurity represents a remarkably underexplored site for this type of research, because risk models – mathematical instruments that translate vulnerabilities and organizational threats from external threats into actionable indices – are among the most transcendental mediating technologies in the contemporary organizational ecosystem. However, its phenomenological dimension has received almost no attention. For this reason, this article addresses this lack by analyzing the Contextual and Multimodal Threat Impact Index (CIIM), which is an original dynamic risk model developed by the author [20,21] for its implementation in undergraduate and graduate education and professional practice in cybersecurity.

The CIIM is suitable for post-phenomenological analysis for three reasons. First, it introduces time projection as a constitutive feature — it calculates not the current state of threat but its predicted evolution at t+1. Second, it treats the condition $R(t) \rightarrow 0$ (where R represents organizational resilience) as a genuine mathematical singularity rather than a numerical limit case that needs to be smoothed out — a design decision with profound phenomenological implications. Third, it incorporates a hybrid machine learning architecture whose multiple components introduce different forms of technological intentionality. We could say that these characteristics make the CIIM not only a technical instrument, but a philosophically rich object for research.

The article is composed of section 1 with the Introduction; Section 2 reviews the theoretical foundations of postphenomenology relevant to our analysis. Section 3 provides a structural analysis of existing cybersecurity risk frameworks from a mediating perspective. Section 4 offers a post-phenomenological reading of the formal architecture of CIIM. Section 5 examines the epistemological implications of the machine learning components of the model. Section 6 develops the concept of 'phenomenology of collapse' as a novel contribution. Article 7 addresses the ethical implications, and finally, Section 8 concludes with a discussion of broader methodological implications for the empirical philosophy of technology.

## 2. Theoretical foundations: post-phenomenology and mediating artifacts

### 2.1 From Husserl to Ihde: The phenomenological inheritance

Postphenomenology inherits from Husserlian phenomenology the fundamental idea that human consciousness is always intentional, because it is always directed towards an object structured by the conditions of its own act of perceiving. For Husserl [8], the task of phenomenology was to describe the essential structures of intentional experience framed in the 'natural attitude', the non-reflexive assumption that the world simply is as it seems. This

Salas-Guerra, R. | 2

method was designed to reveal the constitutive role of consciousness in the construction of experience, a project later extended by Merleau-Ponty [9] to the embodied and perceptual dimensions of lived experience.

On the other hand, Martin Heidegger [6] radicalized this project by insisting that the relevant 'background' conditioning of experience is not consciousness but Being-in-the-world, that is, the practical, embodied, and culturally situated commitment to an environment populated by tools and traditions. For Heidegger, tools in their usual operation become 'transparent' (zuhanden, ready to use): the carpenter's hammer is removed from attention while the carpenter tends to the nail. Only when the tool breaks down, fails or becomes visible does it present itself as an object of theoretical attention (vorhanden, present). This distinction, although originally developed in the context of manual craftsmanship, proves remarkably useful for analyzing digital instruments.

Don Ihde's achievement was to transform this knowledge into a rigorous empirical methodology applicable to the diverse landscape of modern technology. In his seminal study Technics and Praxis [1] and in subsequent works [2,3], Ihde proposed a typology of human-technology relations that has become the standard conceptual vocabulary of postphenomenology. Its four main relationships: incarnation, interpretive, otherness, and background; describe the different ways in which technologies are positioned in relation to the experience of the human world. In the world of embodiment (self-technology) relationships, technology becomes quasi-transparent, embodied in the bodily experience of the self, such as wearing glasses or using a cane. In interpretive relations (I-technological world), technology presents a text or representation that must be read in order to obtain information about the world, such as on a thermometer or a dashboard. In the relations of (I-technology world) otherness, technology itself becomes the quasi-other with which the human interacts. In background relationships, technology operates as an experience that shapes the environment without becoming the object of direct attention.

Peter-Paul Verbeek [4,5] has expanded this framework by developing the concept of 'technological mediation' as a broader category that encompasses not only perceptual relations, but also how technologies measure practical actions, ethical orientations, and the very constitution of subjectivity. For Verbeek, mediating technologies do not transmit a pre-given reality to a pre-determined subject; they coconstitute both poles of the human-world relationship. This 'co-constitution thesis' has significant implications for analysing risk models, as it suggests that such instruments not only measure a pre-existing risk landscape, but actively participate in the construction of what risk means, what counts as a threat and who counts as responsible [22,25].

## 2.2 Technological intentionality

A concept of relevance that must be considered today is 'technological intentionality': that orientation of a technological artifact linked to the world, and incorporated into its design through the choices, assumptions, and values of its creators. Thus, it is established that the artifacts are not neutral conduits; rather, they embody what Langdon Winner [10] called 'politics', but in a sense that they embody phenomenological orientations which direct attention to certain features of the world and away from others.

In risk models, we must state that technological intentionality operates at multiple levels. At the formal level, when the choice of a variable includes how to weight them, it determines which dimensions of the threat ecosystem become visible and salient. On the other hand, at the temporal level, when a model calculates a static snapshot or a dynamic projection, thus determining the organizational security, which appears as a fixed property or as an evolutionary condition. Finally, the threshold level is intended to handle the boundary conditions that occur at the limits of the model's normal operating range, thus determining whether the model can reveal extreme states or only gradual variations.

Consequently, these three levels — formal, temporal, and threshold — are the basis on which the CIIM model is most clearly distinguished from predecessor frameworks, and it is these distinctions that will anchor our post-phenomenological analysis in the later sections of next-generation models.

## 3. Cybersecurity risk frameworks as mediating artifacts

### 3.1 The phenomenological poverty of CVSS

The Common Vulnerability Scoring System (CVSS), currently in version 4.0, is the dominant risk assessment instrument in global cybersecurity practice [11,12]. Managed by FIRST (Forum of Incident Response and Security Teams), this model assigns numerical scores between 0 and 10 to individual vulnerabilities based on metrics such as attack vector, attack complexity, required privileges, user interaction, and impact on confidentiality, integrity, and availability. Therefore, this design is a static and context-independent scoring instrument since the same vulnerability receives the same base score regardless of the specific organizational environment in which it is located.

From a postphenomenological perspective, CVSS establishes a paradigmatic interpretative relationship where the security professional does not encounter the vulnerability itself, but rather a numerical representation which is a text that must be read and that mediates access to the reality of the threat. This alone is not the problem; since all instrumentation implies interpretation. What is phenomenologically significant is the specific character of the interpretative transformation that CVSS performs. Here are three aspects that deserve attention.

First, CVSS operates in the present tense. It marks a vulnerability as it currently exists, with no projection of its likely evolution over time. The temporal experience of the security professional using CVSS is therefore that of a static world interrupted by discrete events as each new vulnerability is assessed in isolation, and the cumulative trajectory of organizational risk remains phenomenologically invisible.

Second, CVSS is context-free at the base score level. The same vulnerability to shock gets, both in a Fortune 500 bank and in a two-person startup. Although the standard provides 'temporal' and 'environmental' metrics as complements, these are inconsistently applied in practice and remain subordinate to the base score in public discourse. Thus, the organizational life world where the specific configuration of assets, dependencies, resilience mechanisms, and operational contexts constitute lived reality is largely excluded from the instrument's field of vision.

Third, and most important for our purposes, CVSS has no mechanism that represents organizational fragility or proximity to systemic failure. A resilient organization and a deeply engaged organization can receive identical CVSS scores for the same vulnerability, even though their existential relationship to that vulnerability is completely different. This is one of the main problems to discuss based on the phenomenological experience of risk because the sense of how close one is to the limit is precisely what CVSS cannot reveal.

## 3.2 The FAIR framework: a probabilistic mediation

The Information Risk Factor Analysis (FAIR) framework [13] represented a significant advance over CVSS in its sophistication of risk modeling. FAIR decomposes risk into Loss Event Frequency (LEF) and Primary Loss Magnitude (PLM), where the LEF is analyzed as a product of Hazard Event Frequency (TEF) and Vulnerability. Unlike CVSS, FAIR is probabilistic: it expresses risk as distributions rather than point estimates, recognizing uncertainty as a constitutive feature of risk assessment.

Phenomenologically speaking, FAIR establishes a different measurement structure. By expressing the results as probability distributions, he introduces what we might call a 'modal dimension' to the analyst's experience in cybersecurity, since the world is not revealed as a set of fixed properties, but as a field of possibilities with associated probabilities. This represents a significant expansion of the experiential horizon of the security professional.

However, it is important to clarify that FAIR shares with CVSS a fundamental limitation from a post-phenomenological perspective, which is that it operates within a framework of actuarial rationality that is oriented only to the quantification of financial losses. In the organizational world, the meanings, values, dependencies, and relationships that constitute the professional environment experienced by the security professional only enter into the model to the extent that they can be translated into monetary terms. Everything else is excluded from the framework.

## 3.3 The mediational gap

What we call 'the mediating gap' in the existing cybersecurity risk frameworks [14–16] that have been discussed is the systematic failure to reveal three dimensions of security expertise that are phenomenologically fundamental to practitioners' understanding, these being the temporal dimension (how risk evolves), the resilience dimension (how close the organization is to collapse), and the contextual dimension (how the organization's specific life world is shapes the importance of the threat). This is precisely the anchor point or gap that CIIM intends to address through arguments that will be explained, this being a phenomenological contribution in the practice of cybersecurity instrumentation.

## 4. A post-phenomenological reading of the CIIM

## 4.1 Formal architecture

CIIM is defined by the following equation:

$$CIIM(t+1) = [A \cdot T(t) \cdot V(t) \cdot E(t)] / R(t) + \alpha \cdot P(t)$$

where A represents the breadth of the threat base (a contextual constant calibrated to the organizational profile), $T(t)$ the active threat level at time t, $V(t)$ the vulnerability index, $E(t)$ the exposure coefficient, $R(t)$ the organizational resilience index, $\alpha$ a weighting coefficient, and $P(t)$ a disturbance function defined as a weighted average of four auditable data sources, such as historical incident data, real-time threat intelligence, user behavior patterns, and detected anomalies.

The model does not produce a score for the current state, but a projection of the threat impact index at t+1, which means the immediately next moment in the modeled time sequence. This forward orientation is the first phenomenologically significant feature of the CIIM design.

## 4.2 Temporal intentionality: The t+1 projection

From a Husserlian perspective [8], temporal experience was never a mere succession of present instants, since each present moment of consciousness contains what Husserl called 'retention' (the just past that is still held in the present) and 'protention' (the anticipated near future that is already implicit in the present), which seeks to establish that authentic temporal experience is structurally anticipatory. In other words, we are always projecting forward into what is about to be.

Most risk models violate this temporal experience structure by presenting only the retained dimension, which is the current score, based on known vulnerabilities and past incidents. In this way, CIIM restores the potential dimension by making the future the main output of its calculation, thus seeking that the security professional who uses CIIM finds himself with a world that is already revealed as being on the way to becoming something like a world with internal momentum, with trajectories, with a direction.

In the end, this has significant practical implications for the professional's agency because when the world is revealed to be static (as is the case with CVSS), the intervention appears as a correction returning an anomalous state to a baseline. When the world is revealed as temporary (as is the case in CIIM), intervention appears as direction redirecting a trajectory already in motion. They are not mere different metaphors; They are different practical orientations that establish two very different forms of professional action and ethical responsibility.

## 4.3 The Resilience Denominator and the Topology of Organizational Space

The most distinctive philosophical feature of the CIIM is the treatment of organizational resilience $R(t)$, a key point in the model, since it acts as the denominator of the primary term. In standard mathematical modeling practice, a denominator that approaches zero is treated as a computational problem that requires a technical solution, since a small epsilon ($\varepsilon$) is added to avoid division by zero, ensuring that the model always returns a finite output, for this reason the CIIM framework explicitly rejects this solution.

In CIIM, $R(t) \to 0$ is treated not as a computational artifact, but as a genuine phenomenological singularity, of a systemic collapse condition in which the standard quantitative relationships between threat variables are completely decomposed. When resilience approaches zero, the organization is not simply at high risk in the ordinary sense; because it has entered a qualitatively different mode of existence in which normal operational continuity is no longer possible. For this reason, the model refuses to smooth out this discontinuity, since doing so would falsify the practitioner's experiential access to organizational reality.

This design decision is epistemologically based on Heidegger's analysis of the rupture in Being and Time [6]. For Heidegger, the familiar structure of practical engagement breaks down when the underlying structure of existence becomes visible—for example, the hammer that breaks reveals the hammer as a hammer and reveals the larger context of the workshop in which the hammer has its meaning. By analogy, the CIIM singularity in $R(t) \to 0$ reveals the organization as an organization, revealing the network of dependencies, relationships, and values that normally remain invisible in the smooth functioning of daily operations.

Phenomenologically, we can describe this as a transformation of what we might call the 'topology of organizational space'. This is because in the normal functioning of the CIIM ($R(t)$ well above zero), the organisational space appears as a fluid and continuous field in which risk can be managed through proportionate interventions. As R(t) approaches zero, this smooth topology transforms and the field becomes more and more curved; producing increasingly unpredictable effects on interventions, where finally, the model indicates that the organization has entered a regime of genuine discontinuity. The analyst who perceives this signal does not receive a high numerical score; as it is undergoing a qualitative transformation in the nature of its organizational world.

### 4.4 The Perturbation Function and the Contextual Basis

The CIIM perturbation function P(t) = f(D_hist, time D_real, B_user, A_patterns) addresses a mediational gap identified in Section 3, which is based on the calculation of the model in four different dimensions of the organizational context. Each data source represents a different temporal and epistemic relationship with organizational reality where historical incident data provides retrospective pattern recognition; this being of transcendental importance in real-time threat intelligence, thus providing contemporary situational awareness; which for this study represents that user behavior patterns provide an index of the internal threat landscape; and the anomalies detected provide early signals of emerging conditions not yet captured by established patterns.

From the perspective of the Ihde typology [1] within the model, P(t) transforms the CIIM from a purely interpretive instrument (presenting a text to read) into something closer to what we might call a 'contextually embodied' instrument, which is partially constituted by the specific organizational lifeworld in which it operates. We should also clarify that the perturbation function ensures that two organizations facing identical threat vectors will nevertheless receive different CIIM projections, because the lived context of each organization contributes significantly to the respective calculation. This represents a very significant step towards what Verbeek [4] would call 'co-constitution': the instrument is not a neutral observer of a predefined risk landscape, but a participant in the continuous constitution of the organizational experience of risk.

## 5. Machine Learning and Distributed Intentionality Components

The equational structure of the CIIM model is complemented by a hybrid machine learning architecture comprising three important components: LSTM/GRU networks [17] for temporal sequence modeling, XGBoost classifiers [18] for risk-level categorization, and a Reinforcement Learning (RL) agent [19] whose reward function is defined as R = -ΔCIIM - λ · action_cost. Each component introduces a distinctive form of technological intentionality into the analyst's perceptual field.

LSTM/GRU networks [17] are temporally oriented because their inductive bias, that is, the structural assumption built into their architecture, is that past sequences predict future states. By deploying these networks as components of CIIM, the model embodies a particular temporal ontology where organizational risk is constituted by its history and its future is always shaped by its trajectory. This is not a neutral assumption; it privileges certain types of organizational change (gradual, historically continuous) and may be less sensitive to discontinuous changes that break with historical patterns.

On the other hand, the XGBoost classifier [18] operates on a different scale of analysis, translating continuous risk indices into discrete categorical levels. This categorical mediation is phenomenologically significant because it transforms a gradient field into a structured set of discrete possibilities that would be High, Medium, Low, Critical, each of which carries specific institutional meanings and triggers specific institutional responses. The analyst who sees

'Critical' is not simply seeing a high number; it is faced with a categorically different organizational situation that activates a pre-established framework of meanings, obligations, and responses. In this scenario it is a clear example of what Verbeek [4] calls 'amplification and reduction': where the classification amplifies certain characteristics of the risk situation (its categorical severity) while reducing others (the precise numerical distance to the thresholds, the uncertainty of the classification, the gradient nature of the underlying risk distribution).

Of all these machine learning components mentioned, one of the most philosophically relevant is the RL agent [19], whose reward function explicitly incorporates cost of action along with risk reduction. This design states that the agent optimizes not only threat mitigation, but also efficiency, as it is trained to recommend interventions that minimize risk while taking into account the organizational costs of the intervention. Phenomenologically speaking, this introduces what we might call 'ethical intentionality' into the architecture of the instrument: the model not only reveals risks, but is already oriented towards a particular normative framework (efficiency, and proportionality) that will shape the interventions it recommends. Therefore, the professional who follows the recommendations of the RL agent does not simply act on the information; It acts within an evaluative framework that has been incorporated into the design of the instrument.

The distributed nature of CIIM's machine learning architecture presents a kind of 'polyphonic intentionality': with multiple partially overlapping perspectives whose convergences and divergences are themselves informative. When LSTM/GRU temporal modeling and XGBoost categorical classification produce consistent signals, the analyst experiences high epistemic confidence. When they diverge, this divergence is itself a sign of organizational complexity that deserves attention.

## 6. Phenomenology of Collapse: A Conceptual Contribution

The above analysis suggested the usefulness of a new concept that is proposed as a contribution to postphenomenological methodology; This is the 'phenomenology of collapse'. Which is activated precisely at the limits of normal instrumental operation and not in the smooth operation of the instrument, but in its decomposition, its saturation or its singularity.

This concept is inspired by Heidegger's analysis of rupture [6], where the breakdown reveals what was previously invisible in a fluid engagement; This analysis is mainly concerned with the phenomenology of the experiencing subject that is now decomposed. The phenomenology of collapse, as developed in this paper, is concerned with the design of instruments that can reveal the approach of collapse before it occurs by means of instruments that, so to speak, are transparent to their own boundary conditions.

Most technical instruments are designed to be robust in their limits because they are designed to fail gracefully, to return sensible outputs even when the conditions for sensible outcomes are no longer met. This robustness is a true technical virtue, but it has a phenomenological cost, the same one that hides from the analyst the information that he most urgently needs, that is, that he is operating within or outside the valid range of the instrument. The epsilon smoothing of $R(t)$ in conventional risk models is a paradigmatic case because this concealment does ensure that the model always returns a finite output, but it does so by hiding the most significant signal possible, which in this case would be the organizational collapse after a simply too high numerical value.

The treatment of $R(t) \to 0$ by the CIIM as a genuine singularity represents an alternative design philosophy that is phenomenologically characterized as responsible, refusing to soften the boundary condition, and instead makes the singularity itself visible as a qualitative discontinuity. This is collapsing phenomenology in practice with the design of an instrument oriented not only towards the normal range of functioning, but towards its own limits, capable of revealing the approach to those limits as a qualitatively different mode of organizational existence.

Therefore, the concept of collapse phenomenology in this work has a broader applicability of cybersecurity risk modeling. Any area in which complex systems can undergo qualitative transitions of state such as ecological systems, financial systems, organizational structures, social movements can benefit from instruments designed under this principle such as instruments transparent to their own singularities, which reveal the collapse approach not only as a very high score, but as a phenomenologically distinct condition of existential fragility.

## 7. Ethical implications: responsibility, agency and the designed gaze

Postphenomenology does not simply describe how technologies measure experience; it also raises normative questions about how it should be mediated. Verbeek's concept of 'moralizing technology' [5], where the idea that technological artifacts embody normative orientations that shape human ethical behavior, is in its essence relevant to the analysis of risk models in cybersecurity.

Every risk model makes implicit ethical decisions, including those that are considered a threat, on whose interests the calculation is represented, and on what trade-offs between security and operational continuity are acceptable. These decisions are usually made by engineers and risk analysts during the design process and then 'frozen' in the architecture of the model, where they operate invisibly on each professional who uses the instrument. Therefore, the phenomenological transparency of these decisions as well as their invisibility in routine usage are in themselves an ethical problem, because it means that professionals may be acting according to normative guidelines that they have not consciously approved.

The inclusion of the cost of action by the CIIM in the RL reward function is a clear example of such a frozen ethical decision. The model has been designed to optimize proportionality; it will not recommend a maximum intervention if that intervention is excessively costly. This is reasonable and defensible normative guidance, but it is guidance that the practitioner may not be aware of at times with a result of non-alignment in their organization's specific normative commitments. An organization with a zero-tolerance policy for data breaches might prefer an instrument that maximizes risk reduction without considering cost, as well as an impact on severe resource constraints, which could welcome the proportionality orientation, but wants it to have a stronger weight.

These observations point to what might be called a 'phenomenological ethics of instrument design': a set of principles for the design of mediating technologies that informs careful attention to how those technologies will shape the practitioner's experience and agency, for this reason, three of these principles are proposed in this paper, informed by our CIIM analysis.

First, the principle of phenomenological transparency where instruments must be designed to make their own mediational structure visible, this allows to reveal, as far as possible, how they are transforming the practitioner's experience rather than presenting themselves as neutral conduits towards a pre-established reality. CIIM partially embodies this principle by including the source distribution of the perturbation function, allowing practitioners to see how each data source contributes to the overall calculation.

Second, the principle of boundary disclosure when instruments must be designed to be transparent in their own boundaries, thus revealing the approach of conditions under which their normal operation is no longer valid, rather than smoothing out these conditions with technically robust, but phenomenologically misleading results. CIIM embodies this principle through its treatment of the uniqueness of resilience.

Thirdly, the principle of ethical legibility where the normative guidelines integrated in the design of an instrument must be explicitly articulated and made available for review by professionals and organisational personalisation. CIIM's modular architecture with its separation of the formal model from the machine learning layer and the perturbation function creates the structural possibility for this type of customization, although its complete realization would require additional interface design work.

## 8. Discussion: Implications for the empirical philosophy of technology

This article holistically follows a specific methodological commitment where it is established that postphenomenological theory, developed mainly in contexts of bodily interaction with physical technologies, can fruitfully be applied to the analysis of abstract and mathematical instruments in contexts of professional practice. The CIIM analysis suggests that this commitment pays off in multiple directions.

For postphenomenological theory, the analysis of cybersecurity risk models expands the scope of empirical research in a productive way, demonstrating that the concept of mediating artifacts is not limited to physical instruments that literally transform sensory experience, but extends to formal instruments that structure the practitioner's conceptual and perceptual access to complex and data-rich environments [24]. It also suggests that Ihde's typology of human-technology relations [1,3] develops principles through the analysis of the use of built-in instruments that may need supplementation when applied to abstract professional instruments. The relationship of the cybersecurity analyst with the CIIM is not any of the four types of Ihde; on the contrary, it combines characteristics of interpretative relations (the model presents a text to be read) with characteristics of relations of otherness (the RL agent functions as a quasi-autonomous recommendation system) and background relations (the temporal orientation of the model structures the analyst's experience even when he does not actively consult it).

For the philosophy of cybersecurity, the central contribution of this article takes force in the argument that the design of risk models is not a technical issue, but a phenomenological and ethical one. The choice of which variables with unprecedented characteristics, how to deal with boundary conditions, how to incorporate machine learning components, and how to present the results to professionals are decisions that shape the professional's experiential access to organizational reality and their capacity for ethical deliberation. For this reason, treating these decisions as merely technical or as questions of precision in modeling rather than experiential design is itself a form of phenomenological naivety that postphenomenological analysis can correct.

For practitioners and instrument designers, practical involvement is a call to what we might describe as 'phenomenologically informed instrument design': a design methodology that starts not only from questions of formal correctness and computational efficiency, but also from questions about how the instrument will shape the practitioner's or analyst's world, what it will make visible and invisible, and what forms of agency and accountability it will allow or exclude. CIIM represents an attempt to practice this methodology; Its design decisions, particularly the treatment of the resilience singularity and the incorporation of contextual disturbances, can serve as reference points for the future development of the instrument.

In this section we will recognize several important limitations of this analysis. First, our phenomenological reading of the CIIM is necessarily interpretative and cannot be validated against the actual experiential reports of the professionals who use the instrument. Empirical research such as mixed studies on how security analysts experience the use of CIIM in comparison to CVSS and other frameworks would be necessary to test and refine the theoretical claims raised here. Second, the concept of collapse phenomenology, while we believe it is theoretically productive, requires further development and application to additional cases before its methodological implications can be fully assessed. Third, the ethical principles proposed in Section 7 are at a relatively high level of abstraction; translating them into specific design guidelines would require interdisciplinary collaboration between philosophers, security engineers, and UX professionals.

These limitations point to a rich agenda for future research at the intersection between postphenomenology, philosophy of science, and cybersecurity studies. We hope this article will serve as a contribution to this emerging field.

## 9. Conclusions

This article has argued formal models of cybersecurity risk in terms of how mediating artifacts in the post-phenomenological sense measure a predefined risk landscape, actively constituting the security professional's experiential access to organizational reality, and shaping the visible and invisible, which demands urgent action in what can be conceptualized with security. Existing frameworks, especially CVSS, embody a form of technological intentionality that systematically conceals three crucial dimensions of the security experience: temporality, organizational resilience, and contextual specificity.

Nonetheless, the CIIM framework, analyzed here as an empirical case study, represents a deliberate intervention in this scientific landscape since its temporal projection (t+1), its treatment of the resilience singularity as a genuine phenomenological discontinuity, and its contextually sensitive perturbation function together constitute a significantly different way of revealing organizational risk which is more faithful to the lived complexity of security practice and more capable of sustaining the forms of temporal, contextual and ethical reasoning that effective cybersecurity requires.

The concept of 'phenomenology of collapse' proposed in Section 6 is the main theoretical contribution of this article, this being a design principle and a mode of dissemination that is activated at the limits of normal instrumental operation, making visible the approach of systemic rupture as a qualitative discontinuity instead of hiding it behind technically robust numerical results. This concept has applicability beyond cybersecurity to any area of monitoring and management of complex systems.

In broader terms, the argument of the article is a call for phenomenological reflection to become a standard component of the design of risk instruments not as a complement to technical development, but as a co-constitutive dimension of it. The instruments through which practitioners to analysts encounter their professional world with an intellectual responsibility and a practical need for the development of technologies that truly serve the forms of understanding, agency, and ethical deliberation that complex organizations require.

## References

1. Ihde, D. *Technics and Praxis: A Philosophy of Technology*; Reidel: Dordrecht, The Netherlands, 1979.

2. Ihde, D. *Technology and the Lifeworld: From Garden to Earth*; Indiana University Press: Bloomington, IN, USA, 1990.

3. Ihde, D. *Postphenomenology and Technoscience: The Peking University Lectures*; State University of New York Press: Albany, NY, USA, 2009.

4. Verbeek, P.-P. *What Things Do: Philosophical Reflections on Technology, Agency, and Design*; Pennsylvania State University Press: University Park, PA, USA, 2005.

5. Verbeek, P.-P. *Moralizing Technology: Understanding and Designing the Morality of Things*; University of Chicago Press: Chicago, IL, USA, 2011.

6. Heidegger, M. *Being and Time*; Macquarrie, J.; Robinson, E., Translators; Harper & Row: New York, NY, USA, 1962. (Original work published 1927)

7. Heidegger, M. *The Question Concerning Technology and Other Essays*; Lovitt, W., Translator; Harper & Row: New York, NY, USA, 1977.

8. Husserl, E. *The Crisis of European Sciences and Transcendental Phenomenology*; Carr, D., Translator; Northwestern University Press: Evanston, IL, USA, 1970. (Original work published 1936)

9. Merleau-Ponty, M. *The Visible and the Invisible*; Lingis, A., Translator; Northwestern University Press: Evanston, IL, USA, 1968.

10. Winner, L. Do artifacts have politics? *Daedalus* **1980**, *109*, 121–136.

11. Mell, P.; Scarfone, K. *A Complete Guide to the Common Vulnerability Scoring System Version 2.0*; National Institute of Standards and Technology: Gaithersburg, MD, USA, 2007. Available online: https://www.first.org/cvss (accessed on 10 April 2026).

12. FIRST. *CVSS v4.0 Specification Document*; Forum of Incident Response and Security Teams: Cary, NC, USA, 2023. Available online: https://www.first.org/cvss/v4.0/specification-document (accessed on 11 April 2026).

13. FAIR Institute. *An Introduction to Factor Analysis of Information Risk (FAIR)*; Risk Management Insight LLC: Columbus, OH, USA, 2020. Available online: https://www.fairinstitute.org (accessed on 13 April 2026).

14. ISO/IEC. *ISO/IEC 27005:2022—Information Security, Cybersecurity and Privacy Protection: Guidance on Managing Information Security Risks*; International Organization for Standardization: Geneva, Switzerland, 2022.

15. NIST. *Risk Management Framework for Information Systems and Organizations*; Special Publication 800-37 Rev. 2; National Institute of Standards and Technology: Gaithersburg, MD, USA, 2018. https://doi.org/10.6028/NIST.SP.800-37r2.

16. MITRE Corporation. *MITRE ATT&CK Framework v14*; MITRE Corporation: Bedford, MA, USA, 2023. Available online: https://attack.mitre.org (accessed on 13 April 2026).

17. Hochreiter, S.; Schmidhuber, J. Long short-term memory. *Neural Comput.* **1997**, *9*, 1735–1780. https://doi.org/10.1162/neco.1997.9.8.1735.

18. Chen, T.; Guestrin, C. XGBoost: A scalable tree boosting system. In *Proceedings of the 22nd ACM SIGKDD International Conference on Knowledge Discovery and Data Mining*, San Francisco, CA, USA, 13–17 August 2016; pp. 785–794. https://doi.org/10.1145/2939672.2939785.

19. Sutton, R.S.; Barto, A.G. *Reinforcement Learning: An Introduction*, 2nd ed.; MIT Press: Cambridge, MA, USA, 2018.

20. Salas-Guerra, R. *CIIM: A Formal Model of Dynamic Risk with Machine Learning for Threat Prediction in Cybersecurity*; Graduate Program in Cybersecurity, Universidad Ana G. Méndez: Gurabo, PR, USA, 2026. Available online: https://ciim.drsalas.us (accessed on 13 April 2026).

21. Salas-Guerra, R. *CIIM Risk Simulator*, Version 2.0 [Interactive web simulator], 2026. Available online: https://ciim.drsalas.us (accessed on 13 April 2026).

22. Rosenberger, R.; Verbeek, P.-P. (Eds.) *Postphenomenological Investigations: Essays on Human-Technology Relations*; Lexington Books: Lanham, MD, USA, 2015.

23. Van den Eede, Y. Tracing the tracker: A postphenomenological inquiry into self-tracking technologies. In *Chasing Technoscience: Matrix for Materiality*; Ihde, D., Selinger, E., Eds.; Indiana University Press: Bloomington, IN, USA, 2011; pp. 143–158.

24. Floridi, L. *The Fourth Revolution: How the Infosphere Is Reshaping Human Reality*; Oxford University Press: Oxford, UK, 2014.

25. Jasanoff, S. The idiom of co-production. In *States of Knowledge: The Co-Production of Science and the Social Order*; Jasanoff, S., Ed.; Routledge: London, UK, 2004; pp. 1–12.